\documentclass[]{aastex631}

\usepackage{amsmath}	% Advanced maths commands
\def\kms {km\,s$^{-1}$}

\def\ch {$\rm CH_3CN$}
\def\so {$\rm SO_2$}
\def\ffas {\hbox{$\,.\!\!^{\prime\prime}$}}

\begin{document}

\title{Infall Motions in the Hot Core Associated with Hyper-Compact HII Region G345.0061+01.794 B}

\correspondingauthor{Toktarkhan Komesh}
\email{toktarkhan.komesh@nu.edu.kz, guido@das.uchile.cl}

\author[0000-0002-3415-4636]{Toktarkhan Komesh}
\affiliation{Energetic Cosmos Laboratory, Nazarbayev University, Astana 010000, Kazakhstan}
\affiliation{Institute of Experimental and Theoretical Physics, Al-Farabi Kazakh National University, Almaty 050040, Kazakhstan}

\author{Guido Garay}
\affiliation{Departamento de Astronomia, Universidad de Chile, Camino el Observatorio 1515, Las Condes, Santiago, Chile}

\author{Christian Henkel}
\affiliation{Max-Planck-Institut für Radioastronomie, Auf dem Hügel 69, 53121 Bonn, Germany}
\affiliation{Xinjiang Astronomical Observatory, Chinese Academy of Sciences, Urumqi 830011, PR China} 
\affiliation{Astronomy Department, King Abdulaziz University, PO Box 80203, Jeddah 21589, Saudi Arabia}

\author{Aruzhan Omar}
\affiliation{Institute of Experimental and Theoretical Physics, Al-Farabi Kazakh National University, Almaty 050040, Kazakhstan}
\affiliation{Faculty of Physics and Technology, Al-Farabi Kazakh National University, Almaty 050040, Kazakhstan}

\author{Robert Estalella}
\affiliation{Departament de Física Quàntica i Astrofísica, Institut de Ciències del Cosmos, Universitat de Barcelona, \\ IEEC-UB, Martí i Franquès 1, 08028 Barcelona, Spain}

\author{Zhandos Assembay}
\affiliation{Institute of Experimental and Theoretical Physics, Al-Farabi Kazakh National University, Almaty 050040, Kazakhstan}
\affiliation{Faculty of Physics and Technology, Al-Farabi Kazakh National University, Almaty 050040, Kazakhstan}

\author{Dalei Li}
\affiliation{Xinjiang Astronomical Observatory, Chinese Academy of Sciences, Urumqi 830011, PR China}

\author{Andr\'es Guzm\'an}
\affiliation{National Astronomical Observatory of Japan, National Institutes of Natural Sciences, 2-21-1 Osawa, Mitaka, Tokyo 181-8588, Japan}

\author{Jarken Esimbek}
\affiliation{Xinjiang Astronomical Observatory, Chinese Academy of Sciences, Urumqi 830011, PR China}

\author{Jiasheng Huang}
\affiliation{Chinese Academy of Sciences South America Center for Astronomy, National Astronomical Observatories, \\Chinese Academy of Sciences, Beijing 100101, PR China}

\author{Yuxin He}
\affiliation{Xinjiang Astronomical Observatory, Chinese Academy of Sciences, Urumqi 830011, PR China}

\author{Nazgul Alimgazinova}
\affiliation{Energetic Cosmos Laboratory, Nazarbayev University, Astana 010000, Kazakhstan}
\affiliation{Institute of Experimental and Theoretical Physics, Al-Farabi Kazakh National University, Almaty 050040, Kazakhstan}
\affiliation{Faculty of Physics and Technology, Al-Farabi Kazakh National University, Almaty 050040, Kazakhstan}

\author{Meiramgul Kyzgarina}
\affiliation{Energetic Cosmos Laboratory, Nazarbayev University, Astana 010000, Kazakhstan}
\affiliation{Institute of Experimental and Theoretical Physics, Al-Farabi Kazakh National University, Almaty 050040, Kazakhstan}
\affiliation{Faculty of Physics and Technology, Al-Farabi Kazakh National University, Almaty 050040, Kazakhstan}

\author{Shukirgaliyev Bekdaulet}
\affiliation{Energetic Cosmos Laboratory, Nazarbayev University, Astana 010000, Kazakhstan}
 \affiliation{Heriot-Watt International Faculty, Zhubanov University, 263 Zhubanov brothers str, 030000 Aktobe, Kazakhstan}
\affiliation{Department of Computation and Data Science, Astana IT University, 55/11 Mangilik El ave, 010000 Astana, Kazakhstan}
 \affiliation{Fesenkov Astrophysical Institute, 23 Observatory str., 050020 Almaty, Kazakhstan}

\author{Nurman Zhumabay}
\affiliation{Energetic Cosmos Laboratory, Nazarbayev University, Astana 010000, Kazakhstan}
\affiliation{Institute of Experimental and Theoretical Physics, Al-Farabi Kazakh National University, Almaty 050040, Kazakhstan}
\affiliation{Institute of Mathematics, Physics and Informatics, Abai Kazakh National Pedagogical University, Almaty 050010, Kazakhstan}

\author{Arailym Manapbayeva}
\affiliation{Institute of Experimental and Theoretical Physics, Al-Farabi Kazakh National University, Almaty 050040, Kazakhstan}
\affiliation{Faculty of Physics and Technology, Al-Farabi Kazakh National University, Almaty 050040, Kazakhstan}
% \collaboration{20}{(AAS Journals Data Editors)}

% \affiliation{}

% \author{Amy Hendrickson}
% \altaffiliation{AASTeX v6+ programmer}
% \affiliation{TeXnology Inc.}

% \author{Julie Steffen}
% \affiliation{AAS Director of Publishing}
% \affiliation{American Astronomical Society \\
% 1667 K Street NW, Suite 800 \\
% Washington, DC 20006, USA}

%% Note that the \and command from previous versions of AASTeX is now
%% depreciated in this version as it is no longer necessary. AASTeX 
%% automatically takes care of all commas and "and"s between authors names.

%% AASTeX 6.31 has the new \collaboration and \nocollaboration commands to
%% provide the collaboration status of a group of authors. These commands 
%% can be used either before or after the list of corresponding authors. The
%% argument for \collaboration is the collaboration identifier. Authors are
%% encouraged to surround collaboration identifiers with ()s. The 
%% \nocollaboration command takes no argument and exists to indicate that
%% the nearby authors are not part of surrounding collaborations.

%% Mark off the abstract in the ``abstract'' environment. 
\begin{abstract}
We report high angular resolution  observations, made with the Atacama Large Millimeter Array in band 6, of high excitation molecular lines of \ch\ and \so\ and of the H29$\alpha$ radio recombination line towards the  G345.0061+01.794 B HC H {\sc ii} region, in order to investigate the physical and kinematical characteristics of its surroundings. Emission was detected in all observed components of the J=14$\rightarrow$13 rotational ladder of \ch\, and in the $30_{4,26}-30_{3,27}$ and $32_{4,28}-32_{3,29}$  lines of \so. The peak of the velocity integrated molecular emission is located $\sim$0\ffas4 northwest of the peak of the continuum emission.
The first-order moment images and channel maps show a velocity gradient, of 1.1 \kms\, arcsec$^{-1}$, across the source, and a distinctive spot of blueshifted emission towards the peak of the zero-order moment. 
The rotational temperature is found 
     to decrease from 252$\pm24$ Kelvin at the peak position to 166$\pm16$ Kelvin at its edge, indicating that our molecular observations are probing a hot molecular core that is internally excited. The emission in the H29$\alpha$ line arises from a region of  0\ffas65 in size, where its peak coincides with that of the dust continuum. 
We model the kinematical characteristics of the "central blue spot" feature as due to infalling motions, suggesting a central mass of 172.8$\pm8.8 M_{\odot}$.
Our observations indicate that this HC H {\sc ii} region is surrounded by a compact structure of hot molecular gas, which is rotating and infalling toward a central mass, that is most likely confining the ionized  region. 
The observed scenario is reminiscent of a
"butterfly pattern" with an approximately edge-on torus
and ionized gas roughly parallel to its rotation axis.

\end{abstract}

%% Keywords should appear after the \end{abstract} command. 
%% The AAS Journals now uses Unified Astronomy Thesaurus concepts:
%% https://astrothesaurus.org
%% You will be asked to selected these concepts during the submission process
%% but this old "keyword" functionality is maintained in case authors want
%% to include these concepts in their preprints.

%% From the front matter, we move on to the body of the paper.
%% Sections are demarcated by \section and \subsection, respectively.
%% Observe the use of the LaTeX \label
%% command after the \subsection to give a symbolic KEY to the
%% subsection for cross-referencing in a \ref command.
%% You can use LaTeX's \ref and \label commands to keep track of
%% cross-references to sections, equations, tables, and figures.
%% That way, if you change the order of any elements, LaTeX will
%% automatically renumber them.
%%
%% We recommend that authors also use the natbib \citep
%% and \citet commands to identify citations.  The citations are
%% tied to the reference list via symbolic KEYs. The KEY corresponds
%% to the KEY in the \bibitem in the reference list below. 

\section{Introduction}

The formation of high-mass stars begins inside dense and massive molecular cores where high-mass protostellar objects accrete at rates between 10$^{-5}$ and 10$^{-3}$ M$_\odot$ yr$^{-1}$ \citep{2014prpl.conf..149T}. These objects finish their Kelvin–Helmholtz (K-H) contraction very rapidly and reach the main sequence \citep{2000A&A...359.1025N,2006ApJ...637..850K}. At this point, the star radiates extreme ultraviolet (UV) photons that ionize its surroundings, producing very small regions of ionized gas, observationally characterized by sizes $\le 0.03$ pc, densities n$_e>$ 10$^{6}$ cm$^{-3}$, and emission measures $>$ 10$^{8}$ pc cm$^{-6}$ \citep{2000RMxAC...9..169K}. These hyper-compact (HC) regions are thought to signpost an early stage of the evolutionary path of High Mass Young Stellar Objects (HMYSOs).

Theoretical calculations show that almost half of the mass of O-type stars is accreted after the K-H contraction and the onset of ionizing radiation \citep{2009ApJ...691..823H,2014ApJ...788..166Z}. How high-mass stars keep accreting despite the onset of the ionizing radiation is not well established. Theoretical works have shown that under steady spherical accretion, radiation pressure inhibits the growth of the protostars. An effective way to circumvent the radiation and ionized gas pressure is accretion from a disk, allowing the accreting material to reach the young high-mass stars much more easily by flowing inward, mainly through the plane perpendicular to the angular momentum vector of the system \citep{1989ApJ...345..464N,2011ApJ...732...20K}. Accretion through a disk may only choke the ionized region near the disk plane, allowing for H {\sc ii} region development in the polar regions \citep{2007ApJ...666..976K}. In this scenario, an HC  H {\sc ii} region should consist of an ionized biconical cavity confined by a rotating and contracting predominantly neutral molecular core.

How does accretion proceed after the onset of the ionizing radiation? How does the envelope material avoid being ionized and blown away by its own pressure? To answer these questions we undertook ALMA Band 6 observations towards a set of luminous embedded HMYSOs associated with HC H{\sc ii} regions in order to simultaneously observe  molecular emission in highly excited transitions of  \so\, and \ch\, and emission from the ionized gas in the H29$\alpha$ hydrogen recombination line. The two molecules have been used to trace velocity gradients, indicative of rotation, towards  hot molecular cores around luminous young high-mass stars in several cases \citep[e.g.,][]{2014A&A...571A..52B,2014ApJ...796..117G,2013A&A...552L..10S}. 
Our  molecular observations are intended to assess whether or not HC H {\sc ii} regions are associated with rotating hot molecular cores  on scales of 3000 AU, as well as to detect inflow motions from the surrounding gas. Our goal is to find evidence of disk accretion and to settle the question as to whether or not accretion onto the HMYSO is maintained after stellar contraction and UV photon injection.  

In this work, we present observations towards the  HC H {\sc ii} region G345.0061+01.794 B \citep[hereafter G345.01 B,][]{2012ApJ...753...51G} associated with IRAS 16533-4009. 
The Spitzer-GLIMPSE survey shows that it is associated with a bright compact mid-infrared source  prominent in the 4.5 $\mu$m band.
The  G345.01 B HC H {\sc ii} region  has the kinematic near and far distances of 1.7 and 14.7 kpc \citep{2007A&A...474..891U}, respectively.
The ambiguity can be resolved through spectrophotometric estimation \citep[e.g.,][]{2020ApJ...904...77G}. In this study, we adopt the spectrophotometric distance of 2.38 kpc \citep{2011MNRAS.411..705M}.
Choosing the far distance not only results in an unrealistically massive cluster but also places it at an impractical distance above the Galactic plane.

 The paper is organized as follows: in \S 2 we describe the observations performed with the Atacama Large Millimeter Array (ALMA); in \S 3 we present the observational results; in \S 4 we discuss the analysis of the data, including the physical relationship between the hot molecular core and the HC H {\sc ii} region; and in \S 5 we present a summary of the main points addressed in this paper.

\section{Observations}

We observed,  using ALMA in Band 6 between 256.3  and 259.6 GHz, the dust continuum and molecular line emission towards the HC H {\sc ii} region G345.01 B (see Table \ref{tab:1}). The observations were carried out, as part of ALMA Cycle 3, during 21 May 2016, using the 12-m array. The ALMA field of view at this wavelength is $\sim22\arcsec$, defined as the FWHM of the primary beam. The phase center of the array was (RA, Dec, J2000) = (16$^h$56$^m$47{$^s$}, --40$^\circ$14$^\prime$25$\arcsec$).
We observed four spectral windows in dual polarization mode. The first window was centered at the frequency of 256.302035 GHz, has a bandwidth of 1875.00 MHz and a spectral resolution of 1.129 MHz. This setup was chosen to map the H29$\alpha$ radio recombination line (RRL) emission from the HC II region.
The second and third windows were centered, respectively, at the frequencies of 258.388716 and 259.599448 GHz each with  234.38 MHz bandwidths and 488.281 kHz ($\sim0.564$ km~s$^{-1}$) channels. These two setups were chosen to observe the emission from the purported hot core in two high excitation temperature lines of \so. The fourth window, centered at the frequency of 257.325000 GHz, has a bandwidth of 468.75 MHz and a resolution of 488.281 kHz. This setup was chosen to observe emission of CH$_3$CN,  a good temperature probe of both the large scale diffuse and small scale dense gas, in the  J=14--13 ladder. \cite{2015MNRAS.447.1826Z} point out that, based on the analysis of IRAS 16547-4247, to detect emission from the inner regions of the rotating core it is important to use molecular transitions with high upper energy levels ($>$ 300 Kelvin). The selected \so\, transitions, $30_{4,26}-30_{3,27}$ and $32_{4,28}-32_{3,29}$, have upper level temperatures of 471 and 531 Kelvin, respectively, and those of the lines in the \ch\, 14--13 ladder range between 92 and 670 Kelvin. 

J1427-4206 was used as bandpass calibrator, J1717-3342 was  observed as phase calibrator, and J1617-5848 was measured as flux calibrators.
Table \ref{tab:1} lists the parameters of each spectral window, the synthesized beams and the rms noise achieved. The integration time on source was 35 minutes. 
Calibration and reduction of these data were done using the Common Astronomy and Software Applications \citep[CASA,][]{2007ASPC..376..127M}.

The continuum was subtracted by selecting the line free channels from the visibilities using the CASA task uvcontsub. These line-free channels in turn were used to create the continuum images directly in tclean.

\begin{figure*}
\centering
	\includegraphics[width=18 cm]{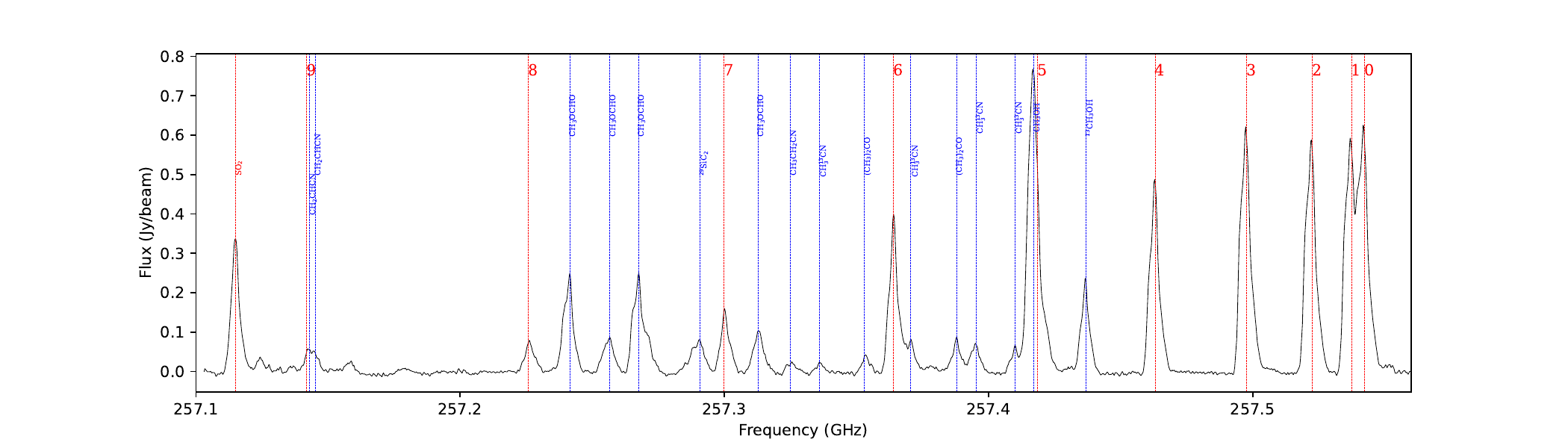}
    \caption{Spectra of the methyl cyanide emission, integrated over a region of 0\ffas5, centered on the G345.01 B HC H {\sc ii} region. K components of the \ch\, 14--13 transition are marked with red dashed lines ($V_\mathrm{LSR}$=--14 \kms). In addition, some other lines detected in this spectral window are marked with blue dashed lines.}
    \label{fig:spec}
\end{figure*}

\begin{figure*}
\centering
	\includegraphics[width=18 cm]{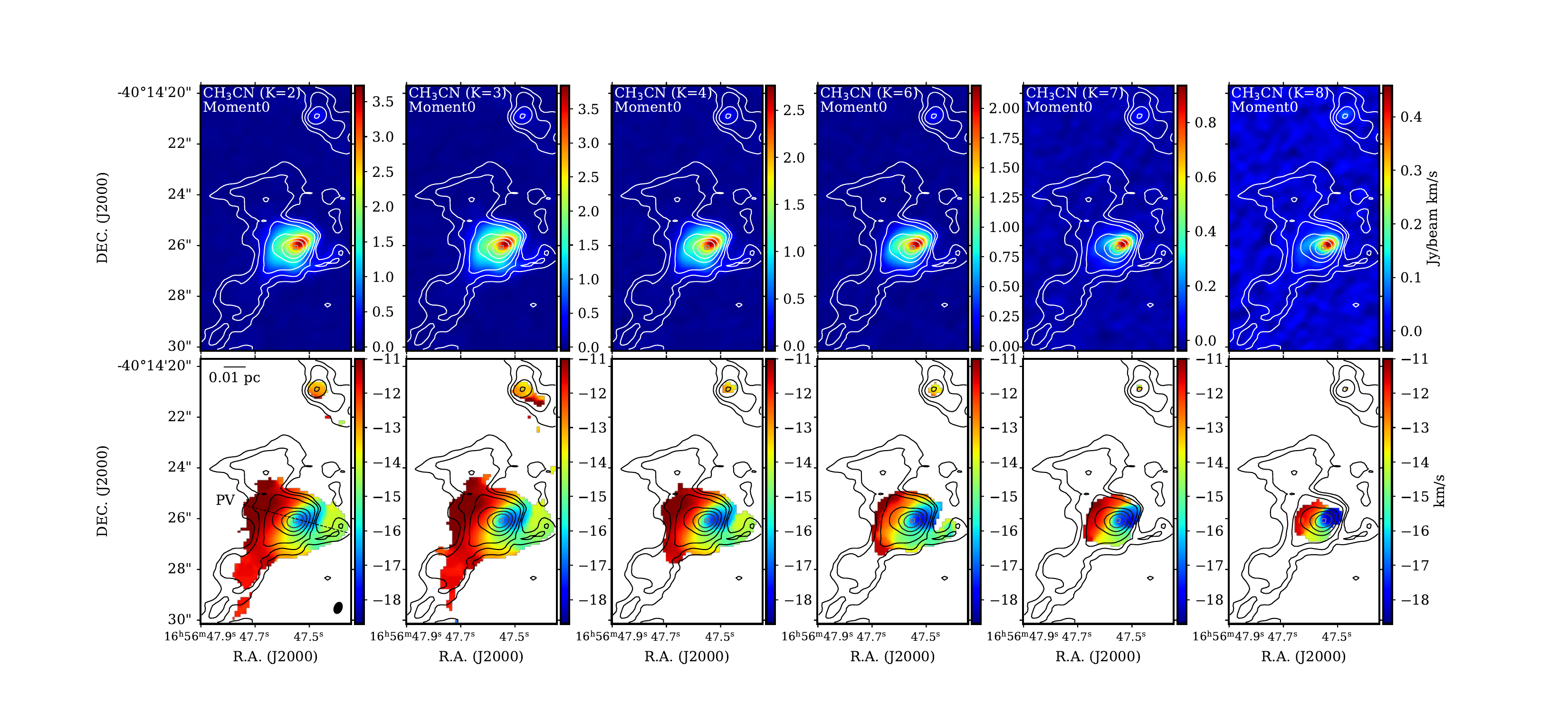}
    \caption{Images of the velocity integrated intensity (moment 0; upper panels) and intensity weighted velocities (moment 1; lower panels) in the \ch\, J=14--13 K=2, 3, 4, 6, 7 and 8 transitions toward the G345.01 B HC H {\sc ii} region. Superimposed are contours of the continuum emission. Contour levels are 10$\sigma$, 20$\sigma$, 40$\sigma$, 100$\sigma$, 200$\sigma$, 400$\sigma$ and 800$\sigma$, where $\sigma$ is 0.3 mJy/beam. 
       The black dashed line shown in the lower left panel indicates the position of the PV cut mentioned in section 3.1.1. 
    The black ellipse shown at the bottom right corner of the lower left panel indicates the beam size.}
    \label{fig:mom345}
\end{figure*}
\begin{figure*}
\centering
	\includegraphics[width=12 cm]{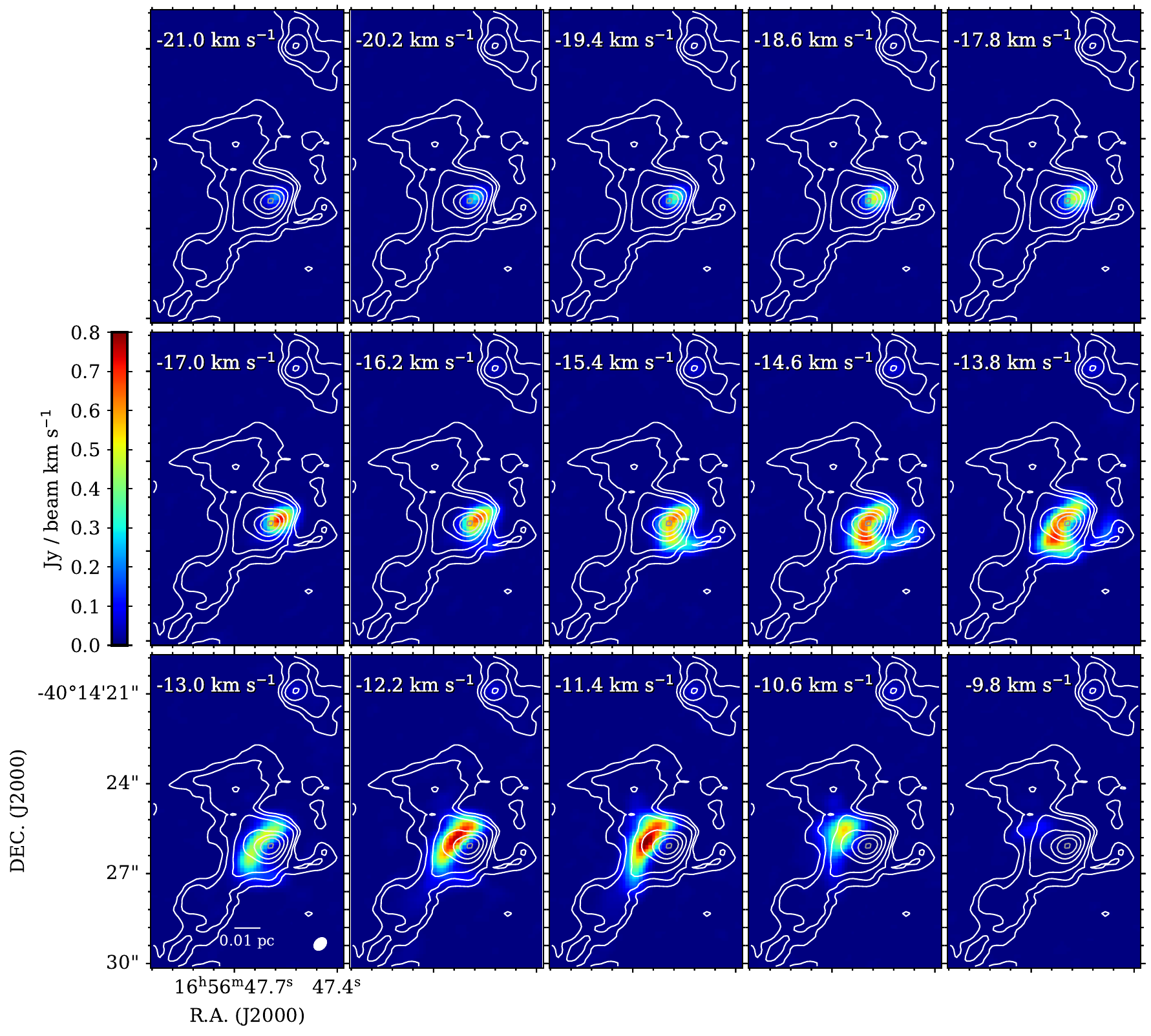}
    \caption{Channel maps for the K=3 component of the \ch\, J=14--13 transition. The contours provide the continuum emission already presented in Fig. 2. The white ellipse shown in the lower right corner of the bottom left panel indicates the beam size.}
    \label{fig:channel}
\end{figure*}

\begin{table*}
	\centering
	\caption{Observational parameters. \label{tab:1}}
	\begin{tabular}{c|cccccc} % four columns, alignment for each
		\hline
		\hline
        Transition & Center Freq. & Bandwidth & Vel. res.&  Synthesized beam & P.A. of the synthesized beams &  rms noise\\
		    & GHz        & GHz     & \kms\,& \arcsec& $^\circ$ & mJy/beam\\
		\hline
		H29$\alpha$                &    256.302035   &   1.875    &    1.320    & 0.52$\times$0.40 &--65.3 &3.8 \\
	\so\, v=0 30(4,26)--30(3,27)   &    259.599448   &   0.23438    &     0.564  & 0.51$\times$0.40 &--64.9 & 4.4 \\
	\so\, v=0 32(4,28)--32(3,29)   &    258.388716   &   0.23438    &     0.566  & 0.51$\times$0.40 & --66.1& 3.9 \\
	\ch\, v=0 14--13 ladder         &    257.325000   &   0.46875    &     0.569 & 0.52$\times$0.40 &--65.2 & 4.0 \\
		\hline
	\end{tabular}
\end{table*}

\begin{table}
	\centering
	\caption Observational parameters of \ch\, J = 14 $\rightarrow$ 13 rotational lines and other lines detected in this spectral window. \label{tab:2}
	\begin{tabular}{c|ccc} % four columns, alignment for each
		\hline
		\hline
		Species & Transition & Frequency & $\rm E_u/k$ \\
		    && (GHz) & (K)  \\
		\hline
  \ch\, &$14_0\rightarrow 13_0$  & 257.527  &     92.70    \\   
	& $14_1\rightarrow 13_1$  &    257.522  &     99.85   \\
	& $14_2\rightarrow 13_2$  &    257.508  &     121.28  \\
	& $14_3\rightarrow 13_3$  &    257.483  &     156.99  \\
	& $14_4\rightarrow 13_4$  &    257.448   &     206.98 \\
	& $14_5\rightarrow 13_5$  &    257.404  &     271.23  \\
	& $14_6\rightarrow 13_6$  &    257.349   &     349.72 \\
	& $14_7\rightarrow 13_7$  &    257.285  &     442.45  \\
	& $14_8\rightarrow 13_8$  &    257.210  &     549.38  \\
	& $14_9\rightarrow 13_9$  &    257.127  &     670.50  \\
        $^{13}$CH$_3$OH & 15( 3,13)-- 15( 2,14) +-- &  257.422 & 321.79 \\
        CH$_3$OH & 18(3,16)--18(2,17) +-- & 257.402 & 446.53 \\
        CH$_3^{13}$CN & 14(1)--13(1), F=14--13& 257.393 & 99.80 \\
        & 14(2)--13(2), F=14--13 & 257.380 & 121.23 \\
        & 14(3)--13(3), F=14--13 & 257.355 & 156.94 \\
        & 14(4)--13(4), F=14--13 & 257.321 & 206.93 \\
        (CH$_3$)$_2$CO & 24(3,22)--23(3,21) EE & 257.373 & 165.68 \\
        & 24(3,22)--23(2,21) AE & 257.338 & 165.75 \\
        CH$_3$CH$_2$CN & 30(0.30)--29(0,29) & 257.310 & 193.54 \\
        CH$_3$OCHO & 21(9,13)--20(9,12)A & 257.298 & 377.19 \\
         &  23(2,22)--22(2,21)E & 257.242 & 342.23 \\
         & 20(5,15)--19(5,14)E & 257.227 &142.79 \\
         CH$_3$OCHO &  20( 5,15)--  19( 5,14)A &257.253 &142.79 \\
         $^{29}$SiC$_2$ & 11(4,8)--10(4,7) & 257.276 & 105.23 \\
         SO$_2$ & 7(3,5)-- 7(2,6) & 257.100 & 47.84 \\
         
		\hline
	\end{tabular}
\end{table}

\section{Results}

\subsection{Molecular emission}\label{sec:3.1}

\subsubsection{\ch\,}

Figure \ref{fig:spec} presents the spectrum of the emission in the J=14$\rightarrow$13 rotational transition of \ch\, integrated over a region of 0\ffas5 in diameter, centered on the G345.01 B HC H {\sc ii} region. Other lines are also present. The 
         stronger ones are indicated in Fig. 1 and briefly addressed 
         in Sect. \ref{sec:infall}. Their transitions, line frequencies and upper state energy levels are given in Table \ref{tab:2}. The J=14$\rightarrow$13 rotational transition of \ch\, consists of 14 $K$ components ($K$=0, 1, ...13; $K$ being the projection of the total angular momentum of the molecule onto its principal rotation axis) of which ten lie within the observed spectral window (red dotted lines).

Figure \ref{fig:mom345} displays images of the zero-order (upper panels) and first-order moments (lower panels) of the emission in the K=2, 3, 4, 6, 7 and 8 components of  the J=14--13 ladder of \ch. Moments of the K=0, 1, 5, 9 components are not shown since they are blended with each other or with other molecular lines (see Fig. \ref{fig:spec}).
Superimposed are contours of the continuum emission. The peak of the velocity integrated intensity \ch\, emission is located $\sim$0\ffas4$\pm$0\ffas1 northwest of the peak of the continuum emission. The first-order moment images  show a velocity gradient from roughly east to west with average velocities
preferentially blueshifted on the western side and redshifted on the eastern side. A spot of blueshifted emission is seen towards the peak of the zero-order moment. The "blue spot" feature is present in all K components shown in Figure \ref{fig:mom345}, confirming that its detection is a robust result.

Fig. \ref{fig:channel} presents channel maps of the emission in the K=3 component, which clearly exhibits the shift in velocity, from 
blueshifted velocities in the West to redshifted velocities in the East. 
 Additionally, a "butterfly pattern" is also noteworthy, which may indicate the presence of accretion disks around high-mass stars.

\begin{figure}
\centering
	\includegraphics[width=0.5\textwidth]{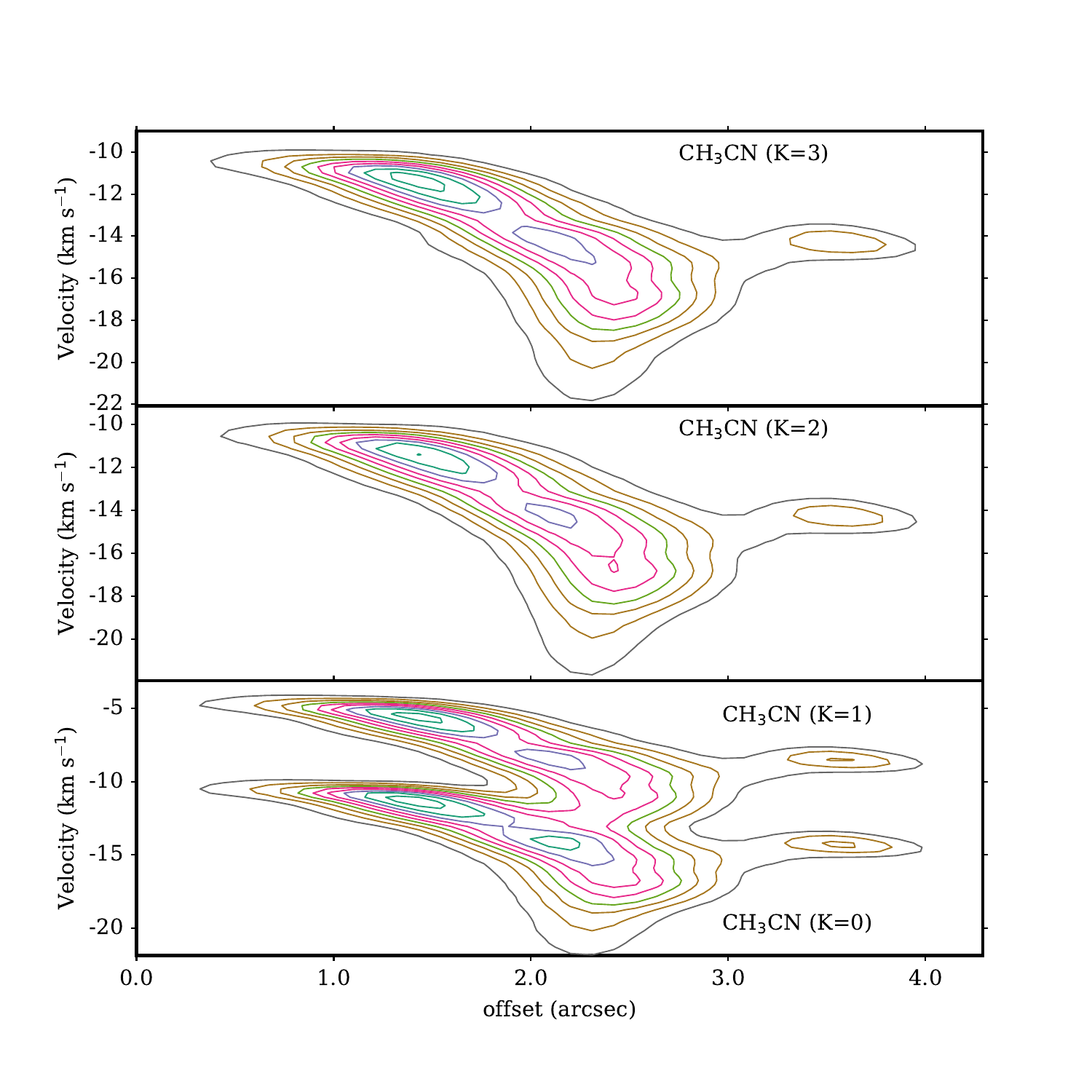}
    \caption{Position-velocity diagrams of K=0, 1, 2 and 3 components of the \ch\, J=14--13 transition along a direction from the red  to the blue part, with a position angle (P.A.) of 255\textdegree\, passing through the continuum peak (see Fig.\ref{fig:mom345}). Contour levels are 10\%, 20\%, 30\%, 40\%, 50\%, 60\%, 70\%, 80\% and 90\% of the peak value of 0.7 Jy/beam.}
    \label{fig:pv}
\end{figure}

The "blue spot" feature and a velocity change are illustrated in Fig. \ref{fig:pv}, which presents position-velocity diagrams of the emission in the $K$=0, 1, 2, 3 components along a direction from the red part to the blue part, with a position angle (P.A.) of 255\textdegree\, passing through the continuum peak.
   There is a clear change in velocity across the source of 4.3 \kms\ over a region of 3\ffas8 (equivalent to 98 \kms\ pc$^{-1}$ at a distance of 2.38 kpc). If this velocity gradient is due to gravitationally bound rotation, it implies a dynamical mass within a 0.023 pc radius of 51 $M_{\odot}$, less than half the mass of the central object as derived in Sec. \ref{sec:infall}.  We conclude that the hot molecular gas is likely bound and rotating around the central object.

\subsubsection{\so\,}

In addition to the high excitation lines of \so\  observed on purpose in this work, the spectral window of the RRL encompasses four  additional transitions of \so\ all of which connect levels with less than 50\,Kelvin above the ground state.
  The transitions and their parameters are listed in Table \ref{tab:SO2-RD-parameters}; col.(2) gives the frequency, col.(3) the energy of the upper state, col.(4) the Einstein A coefficient,  and col.(5) the statistical weight of the upper state. 

\begin{figure*}
\centering
	\includegraphics[width=18 cm]{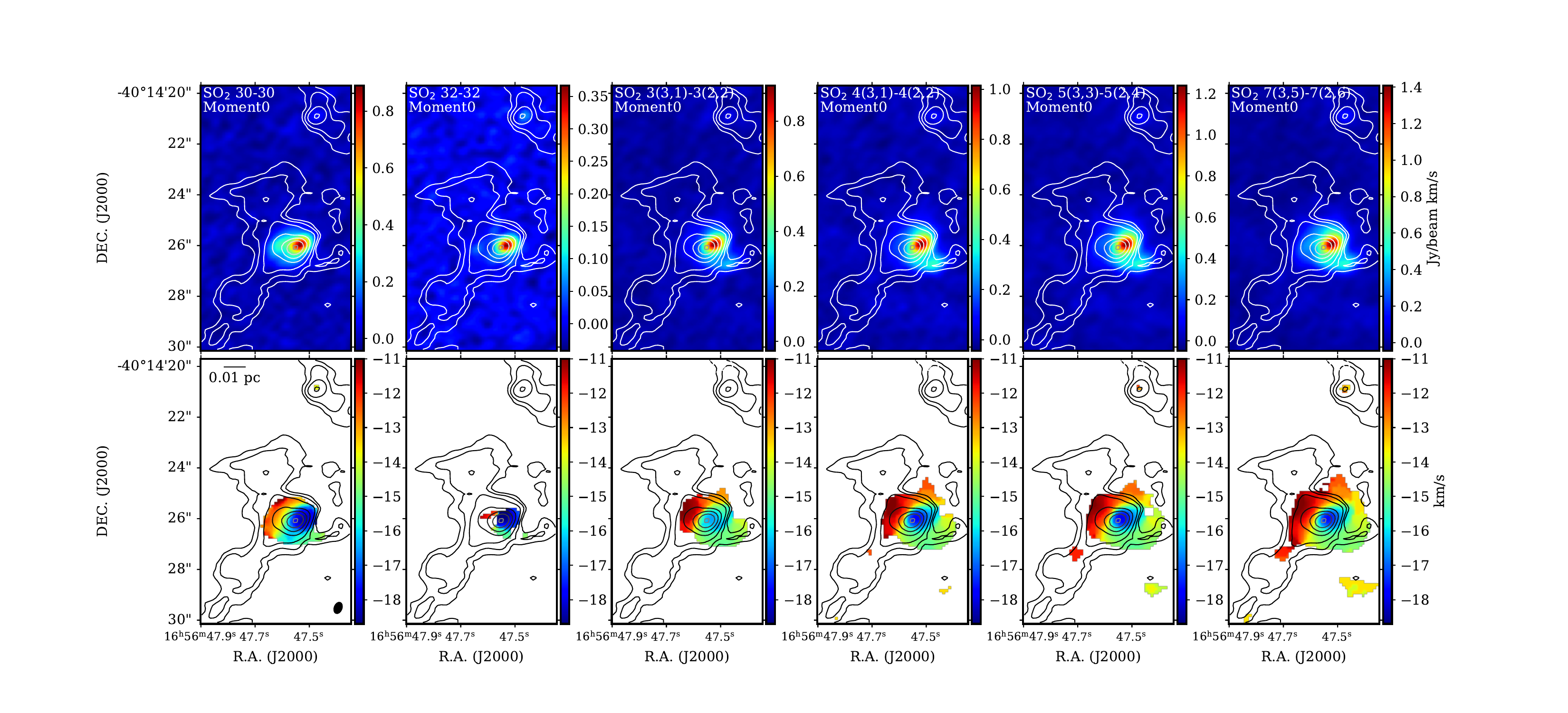}
	\caption{Same as Fig. \ref{fig:mom345} for $30_{4,26}-30_{3,27}$, $32_{4,28}-32_{3,29}, 3_{3,1}-3_{2,2}, 4_{3,1}-4_{2,2}, 5_{3,3}-5_{2,4}$ and $7_{3,5}-7_{2,6}$ \so\ emissions}
		\label{fig:so2}
\end{figure*}

Figure \ref{fig:so2} shows images of the velocity integrated intensity (upper panels) and intensity-weighted velocity (moment 1; lower panels) in all six observed \so\ lines, in order of increasing excitation temperature. 
The peak position of the integrated intensity in \so\, is similar ($<$0\ffas01) to that in the \ch\ lines.
The "blue spot" signature is also present in the \so\, moment one maps.

\begin{table}
\centering
\caption{Parameters of the observed SO$_2$ transitions. \label{tab:SO2-RD-parameters}}
\begin{tabular}{ccccc}
\hline
\hline
Transitions & Frequency & $\rm E_u/k$  & $A_\mathrm{ul}$ & $g_\mathrm{u}$\\
 & GHz & Kelvin & 10$^{-4} $cm$^{-1}$ &\\
\hline
\multicolumn{2}{l}{\emph{High excitation lines}}&&&\\
$30_{4,26}-30_{3,27}$& 259.599 & 471.50 & 2.07 & 61 \\
$32_{4,28}-32_{3,29}$& 258.389 & 531.10 & 2.1 & 65 \\
\multicolumn{2}{l}{\emph{Low excitation lines}}&&& \\
$3_{3,1}-3_{2,2}$ & 255.958 & 27.62 & 0.66 & 7 \\
$4_{3,1}-4_{2,2}$ & 255.553 & 31.29 & 0.93 & 9   \\
$5_{3,3}-5_{2,4}$ & 256.247 & 35.89 & 1.07 & 11 \\
$7_{3,5}-7_{2,6}$ & 257.100 & 47.84 & 1.22 & 15  \\
\hline
\end{tabular}
\end{table}

\begin{figure}
\centering
	\includegraphics[width=\textwidth]{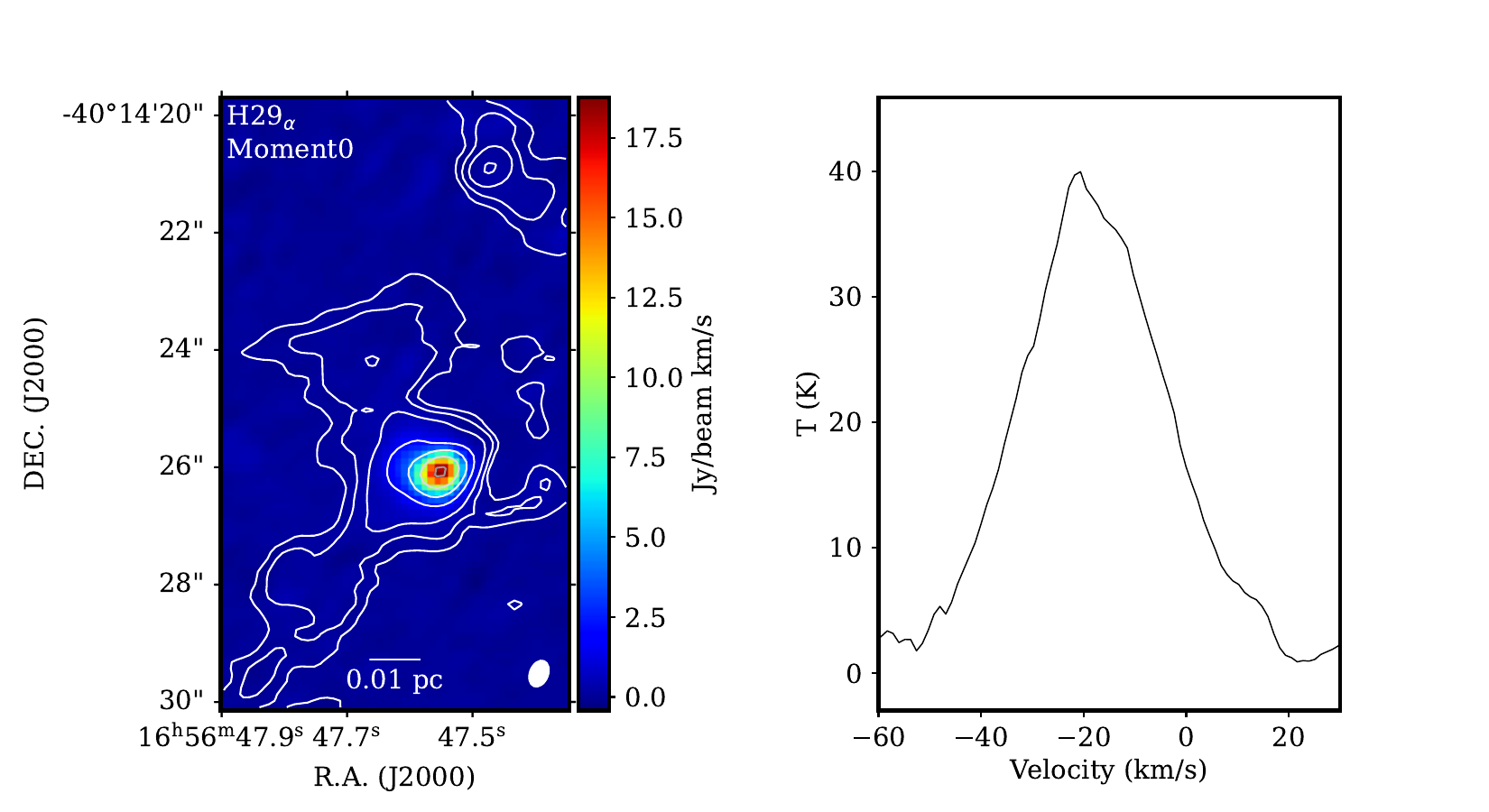}
	\caption{Left panel: an image of the velocity integrated H29$\alpha$ RRL emission along with the dust continuum contours (see Fig.\ref{fig:mom345}). Right panel: a spectrum of the H29$\alpha$ RRL emission integrated over the source. A Gaussian fit to the line profile gives a linewidth of 33.7$\pm$2.3 \kms\, and a line center velocity of $V_\mathrm{LSR}$=--18.1$\pm$0.9 \kms. }
		\label{fig:Ha}
\end{figure}

\begin{figure*}
\centering
	\includegraphics[width=15 cm]{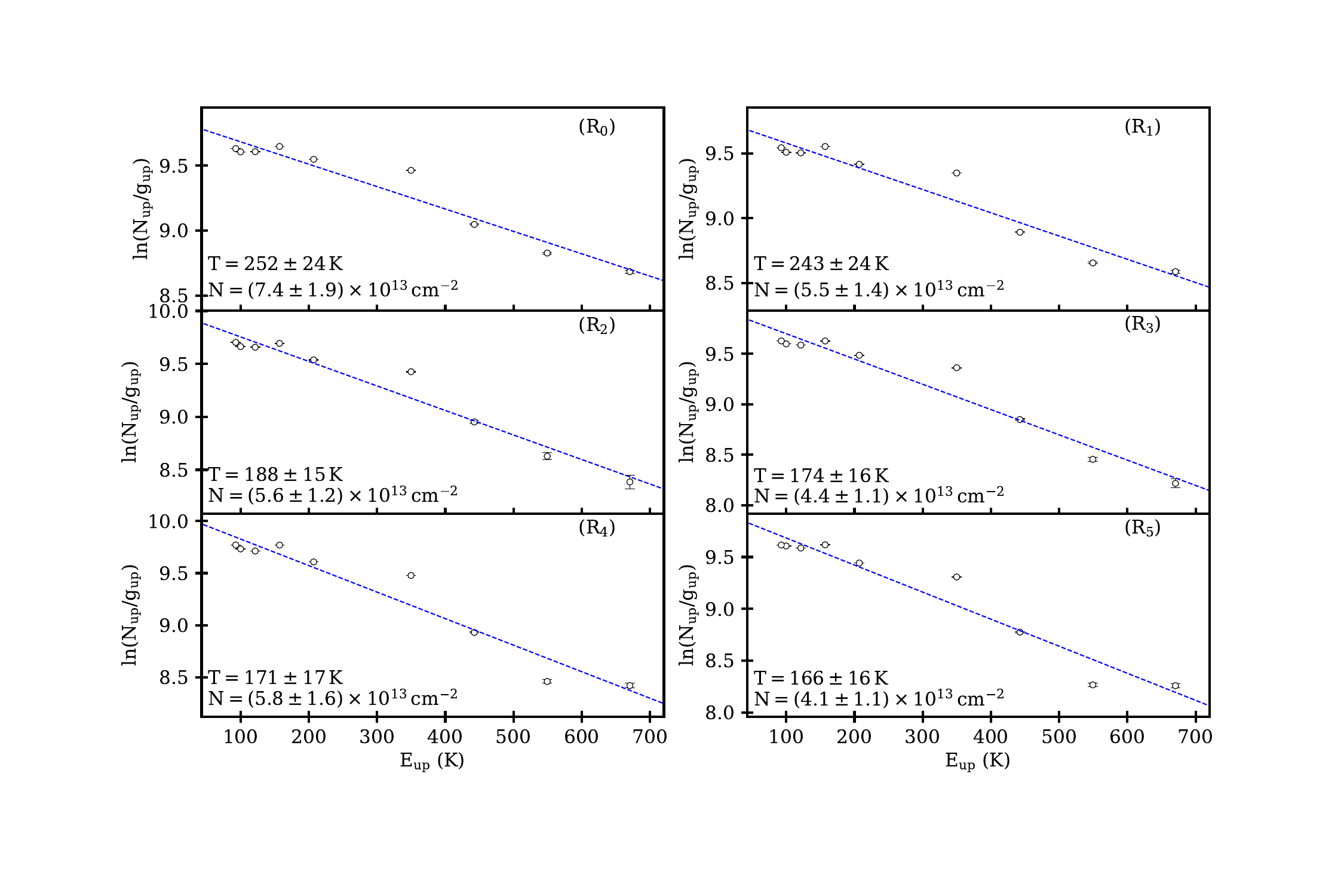}
	\caption{ Rotational diagrams  of \ch\, derived from the peak position ("blue spot", in lower panels of Fig. \ref{fig:mom345}), and in half-rings going from the peak position towards the southeast (position angle: 135$^{\circ}$) with different radii (R$_1$ to R$_5$) from the peak position  using \texttt{MADCUBA}. The rings are centered at the peak position and have widths of 0\ffas15. The inner ring (R1) has an inner radius of 0\ffas33. The rotational temperature decreases from the peak position to the edge of the source ($\sim$1\arcsec) from 252$\pm24$ to 166$\pm16$ Kelvin. }
		\label{fig:RD}
\end{figure*}

\subsection{Ionized gas: H29$\alpha$ RRL emission \label{sec:H29a}}

Since at Band 6 frequencies the continuum is likely to be dominated by dust emission, hydrogen recombination lines (HRLs) become the most direct way to trace the ionized gas. Figure \ref{fig:Ha} (left panel) shows an image of the velocity integrated H29$\alpha$ emission along with the dust continuum contours. Intensities were integrated from $V_\mathrm{LSR}$=--44 to 8 km$~$s$^{-1}$. The position of the peak in the velocity integrated line emission,  of 18.7 Jy/beam km$~$s$^{-1}$, is coincident with that of the dust continuum (RA, Dec) (J2000) = (16$^h$56$^m$47.6$\pm0.1${$^s$}, --40$^\circ$14$^\prime$26\ffas07$\pm{0.10}$).

A Gaussian fit to the observed H29$\alpha$ brightness distribution indicates that the HC HII region has a deconvolved  angular size (FWHM) $\theta_s=\sqrt{0\ffas75\times0\ffas56}\approx0\ffas65$, corresponding to the geometrical mean of the deconvolved major and minor axes.  At the distance of 2.38 kpc this implies a diameter of 0.0075 pc.

Figure \ref{fig:Ha}, right panel, shows a spectrum of the H29$\alpha$ RRL emission integrated over the source. A Gaussian fit to the line profile gives a linewidth of 33.7$\pm$2.3 \kms\, and a line center velocity of $V_\mathrm{LSR}$=--18.1$\pm$0.9 \kms. 
 This velocity differs from those of the \ch\, and \so\, molecular lines. It may be due to the fact that ionized gas and molecular cloud are slightly displaced.

The electron temperature $T^{*}_e$ can be derived from the following expression \citep{2002ASSL..282.....G} assuming local thermodynamic equilibrium (LTE) conditions:

\begin{equation}
	\begin{split}
T^*_e=\Bigg[\bigg(\frac{6985}{\alpha(v,T_e)}\bigg)\bigg(\frac{\Delta V_{\mathrm{H29\alpha}}}{\mathrm{km\,s}^{-1}}\bigg)^{-1} \bigg(\frac{S_\mathrm{ff}}{S_\mathrm{H29\alpha}}\bigg) \bigg(\frac{\nu}{\mathrm{GHz}}\bigg)^{1.1} \\
\times \bigg(1+\frac{N(\mathrm{He^+})}{N(\mathrm{H^+})}\bigg)^{-1}\Bigg]^{0.87},
    \end{split}
\end{equation}
where $S_\mathrm{ff}$ is the free-free continuum flux density, ${S_\mathrm{H29\alpha}}$ is the H29$\alpha$ peak flux density, 
$\alpha(v,T_e)$ $\sim$ 1 is a slowly varying function of frequency \citep{1967ApJ...147..471M}, and $N(\mathrm{He^+})/N(\mathrm{H^+})$ is the $He^+$ to $H^+$ abundance ratio. The free-free flux density, $S_\mathrm{ff}$, cannot be derived from the continuum emission at 256 GHz  because of the contribution of dust emission at this frequency.  We estimate it using the parameters of the HC HII region ({Emission Measure (EM)} and size) derived by \cite{2012ApJ...753...51G}, from a fit to the observed radio continuum spectra at lower frequencies,  obtaining  a value of $S_\mathrm{ff}$=325$\pm$65 mJy.
Using the observed values of ${S_\mathrm{H29\alpha}}$=883$\pm$47  mJy and $\Delta V_{\mathrm{H29\alpha}}$=33.7$\pm$2.3 \kms\, and adopting a value of 0.096 for the $He^+$ to $H^+$ abundance ratio \citep{1994ApJS...91..713M},
we get an electron temperature $T^*_e$=8094$\pm$1534 Kelvin.

Further parameters of the region of ionized gas can be computed using the equations presented in \cite{1967ApJ...147..471M} and \cite{1974A&A....32..269M}. Assuming that the HII region is spherical and homogeneous, using the values of the continuum flux density at 256 GHz (325$\pm$65 mJy), the angular size (0\ffas65), electron temperature (8094$\pm$1534 Kelvin) and distance (2.38 kpc), we determine an electron density of (2.1$\pm$0.2$)\times$10$^5$ cm$^{-3}$, an emission measure of (4.6$\pm$0.9)$\times$10$^8$ pc cm$^{-6}$, a mass of ionized gas (3.6$\pm$0.3)$\times$10$^{-3}$ $M_{\odot}$  and the number of ionizing photons required to excite the HII region becomes (2.6$\pm$0.6$)\times$10$^{47}$s$^{-1}$. 
The errors do neither include the distance uncertainty nor potentially strong density gradients as suggested by \cite{2000prpl.conf..299K}. If we consider the near and far distances of 1.7 and 14.7 kpc, the inferred Lyman continuum flux suggests the presence of a massive star equivalent to a zero-age main-sequence star of type B0 and O5.5, respectively, within the G345.01 B HC H II region \citep{1973AJ.....78..929P} which corresponds to a stellar object with a mass of 10 $M_{\odot}$ to 30 $M_{\odot}$.
% This finding aligns with the results of \cite{2012ApJ...753...51G}, who utilized the near distance and associated it with one of the three infrared sources reported by \cite{2011A&A...525A.149M} within the region. 

\subsection{Dust continuum}\label{sec:continuum}

Continuum contours are overlaid on all moment maps depicting \ch\,, \so, and H29$\alpha$ RRL in Figs. \ref{fig:mom345}, \ref{fig:channel}, \ref{fig:so2}, and \ref{fig:Ha}a. Continuum contour levels are set at 10$\sigma$, 20$\sigma$, 40$\sigma$, 100$\sigma$, 200$\sigma$, 400$\sigma$ and 800$\sigma$, with $\sigma$ being 0.3 mJy/beam. The peak position, measuring 0.234 Jy/beam, aligns ($<$0\ffas01) with that of the ionized gas (H29$\alpha$ RRL), while the peak of the velocity-integrated intensity of \ch\, and \so\, emissions is situated (as mentioned before) approximately 0\ffas4 northwest of the continuum emission peak.
We estimate the dust continuum flux of 565$\pm$112 mJy by subtracting the contribution of the free-free emission of 325$\pm$65 mJy from the total flux of the continuum map of 890$\pm$47 mJy.

The mass of the core (central star) is obtained from the following equation using the dust continuum flux:
\begin{equation}
M_{\text{dust}} = \frac{S_\mathrm{1.1mm} \times D^2}{k_\mathrm{1.1mm} \times B_\mathrm{1.1mm}(T_\text{dust})},
\end{equation}
where $S_\text{1.1mm}$ is the dust continuum flux, $D$ is the distance to the object, and $T_\text{1.1mm}$ is the dust temperature. $B_\mathrm{1.1mm}(T_\text{dust})$ represents the Planck function (black body radiation) at the dust temperature $T_\text{dust}$, and we adopt a grain emissivity, $k_\mathrm{1.1mm}$ = 0.0078 cm$^2$ g$^{-1}$, which is the value calculated by \cite{1994A&A...291..943O} for bare grains and dense gas, where a gas-to-dust ratio of 100 is assumed. Adopting $T_\text{dust}$ = 250 K (see Section \ref{sec:T_rot}), we derive a dust mass of $\sim$4 M$_{\odot}$ (for a distance of 2.38 kpc). This dust mass is significantly smaller than the mass of the central object as derived in Sec. \ref{sec:infall}, while its gas mass is larger.
We conclude that the gravitational field is significantly influenced by the central star.

\section{Discussion}
\subsection{Rotational temperature of \ch}\label{sec:T_rot}
Rotational temperature and total column density of \ch\,were derived from the MAdrid Data CUBe Analysis \citep[\texttt{MADCUBA}\footnote{\texttt{MADCUBA} is a software developed in the Madrid Center of Astrobiology (INTA-CSIC) which enables to visualise and analyse single
spectra and data cubes: https://cab.inta-csic.es/madcuba/.},][]{2019A&A...631A.159M} software assuming LTE.

\begin{figure}
\centering
	\includegraphics[width=0.5\textwidth]{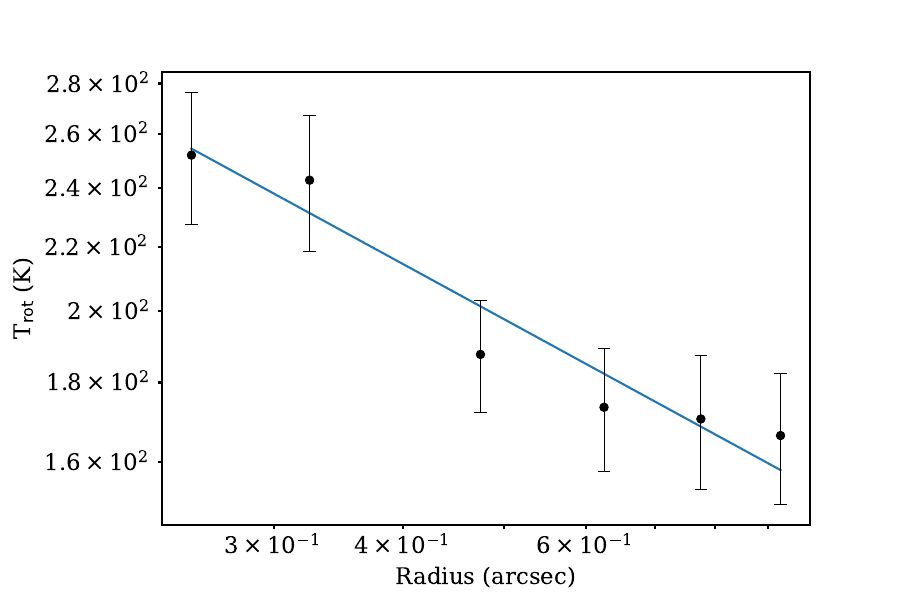}
	\caption{Rotational temperature of the \ch\, versus projected distance from the "blue spot" (see the lower panels in Fig. \ref{fig:mom345}). A power law fit to the observed dependence gives $T_\mathrm{rot} = 154\,r^{-0.35\pm0.19}$.}
		\label{fig:T_rot_vs_dis}
\end{figure}

Figure \ref{fig:RD} displays rotational diagrams of the emission at the peak position ("blue spot" in Fig. \ref{fig:mom345}) and in half-rings going from the peak position towards the South-East (diameters at position angles of 135\textdegree) with different radii (R$_1$ to R$_5$) from the peak position. The rings are centered at the peak position and have widths of 0\ffas15. The inner ring (R1) has an inner radius of 0\ffas33.  The rotational temperature decreases outwards with distance from the "blue spot" (exciting star) being 252$\pm24$ Kelvin at the peak position and  166$\pm16$ Kelvin at the edge of the molecular (\ch) structure ($\sim$0.01 pc). Figure \ref{fig:T_rot_vs_dis} plots the computed rotational temperature versus the projected distance from the "blue spot". A power law fit to the observed dependence gives $T_\mathrm{rot} = 154\,r^{-0.35\pm0.19}$.
This result suggests that the molecular gas is heated via collisional excitation with hot dust, which in turn is heated by the absorption of radiation emitted by the central star \citep{1976ApJ...206..718S}. Using expression (11) in \citet{1999PASP..111.1049G}, we infer that the power-law index of dust emissivity at far infrared wavelengths,  $\beta$, is 0.55 with an emmisivity, $f$, of 0.08 and that the luminosity of the central object is 1.1$\times$10$^4$ L$_{\odot}$. This luminosity is in good agreement with the luminosity of the star ionizing the HC HII region as determined by \cite{2012ApJ...753...51G} from radio continuum observations.

\subsection{Rotational temperature of \so}

\begin{figure}
\centering
   \includegraphics[width=0.5\textwidth]{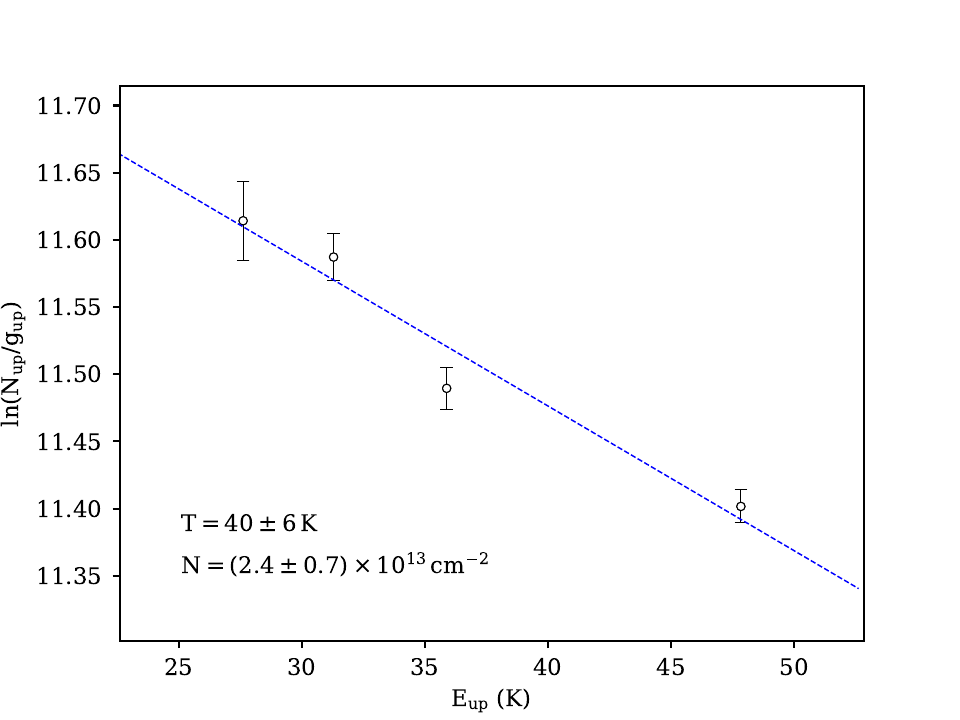}
    \caption{Rotational diagram using the emission from the low excitation lines of \so\,integrated over the whole source. A linear fit to the observed trend implies an excitation temperature of 40$\pm$6 Kelvin.}
    \label{fig:RD_so2}
\end{figure}

From the emission of the four low excitation lines of \so\, shown in Table \ref{tab:SO2-RD-parameters} we obtain the rotational temperature of the envelope  with \texttt{MADCUBA}. 
Figure \ref{fig:RD_so2} plots a rotational diagram of emission from the whole region.
A linear fit to the observed trend implies a temperature of 40$\pm$6 Kelvin. 
Since the data are well matched by a linear
     fit, all \so\, lines appear to be optically thin.

\begin{figure*}
\centering
   \includegraphics[width=\textwidth]{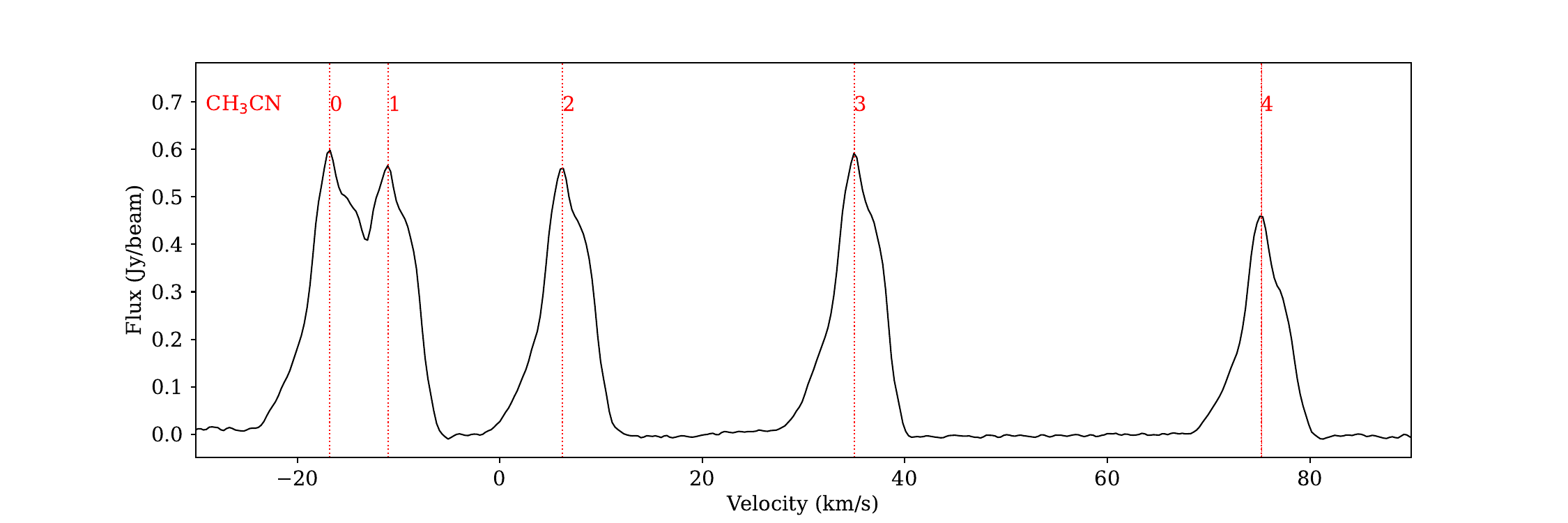}
    \caption{Observed line profiles of several K components of the \ch\, J=14--13 transition (marked with red dashed lines) toward the "blue spot" position (the angular size is $\sim0$\ffas4, V$_\mathrm{LSR}$=--17 \kms).}
    \label{fig:bs_spec}
\end{figure*}

\begin{figure*}
\centering
\includegraphics[width=10 cm]{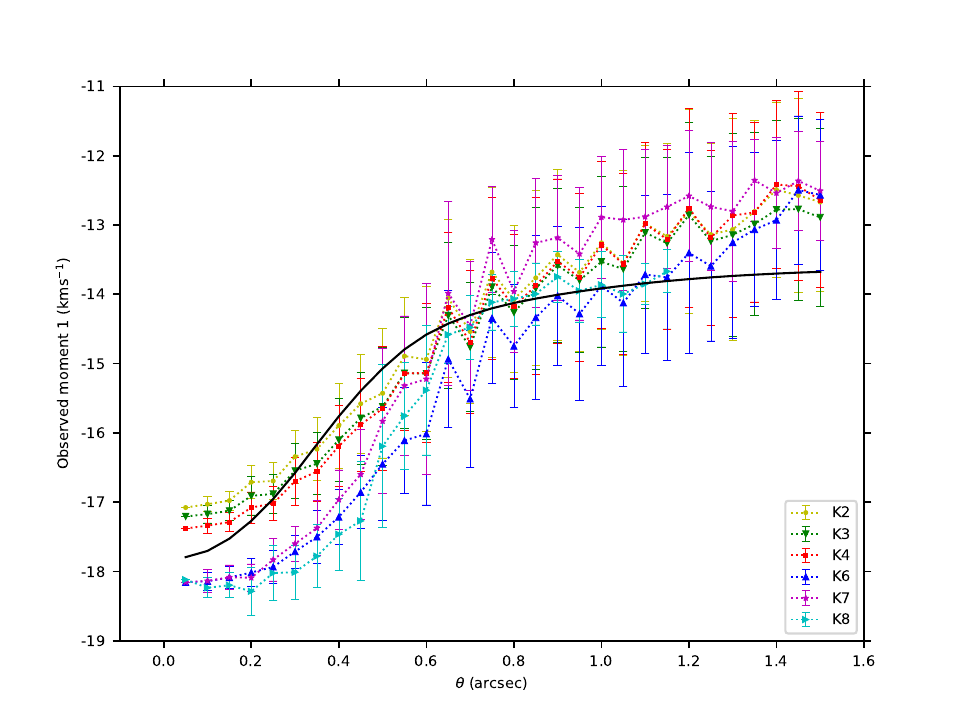}
	\caption{G345.01 first-order moment for the \ch\, transitions as a function of angular distance, measured for full rings of width 0\ffas05 centered on the average position of the peak of the "blue spot", (RA, Dec) (J2000) = (16$^h$56$^m$47.6{$^s$}, --40$^\circ$14$^\prime$25\ffas98), and average radius $\theta$. The error bars are the standard deviation of uncertainties of the velocity inside each ring.  The black solid line shows the best fit to the data. The best fit is obtained for an infall radius much larger than the beam size, an ambient gas velocity $V_\mathrm{LSR}$=--12.49$\pm$0.16 \kms, and a central mass of 172.8$\pm8.8 M_{\odot}$ (for a distance of 2.38 kpc). }
		\label{fig:infal}
\end{figure*}

\subsection{Infall motions}\label{sec:infall}
Within the spectral window covered by the 
         ALMA data (see Fig. \ref{fig:spec}), additional spectral features are 
         also detected. Not only the line profiles of \ch\, and \so\,
         but also the lines of other species such as CH$_3$OCHO and
         $^{13}$CH$_3$OH, integrated over a region of 0\ffas5 in size and 
         centered on the G345.01 B HC HII region, show a blue 
         asymmetry in their line profiles. 
Fig. \ref{fig:bs_spec} shows the observed line profiles of several K components of the \ch\, J=14--13 transition (marked with red dashed lines) toward the "blue spot" position. These spectra also show a blue asymmetry in their line profiles. The appearance of the line profiiles is consistent with the detection of the "classic" asymmetric line profile infall signature \citep{1998ApJ...496..292N}.
Furthermore, \cite{2014MNRAS.437.3766M} and \cite{2019A&A...626A..84E} suggested that the "blue spot" feature mentioned above (see §\ref{sec:3.1}) is a clear signature of infall.
The observation of a blue spot, coupled with the absence of a corresponding red spot, aligns well with the proposed infall scenario. This discrepancy can be attributed to the obscuration of the putative red spot, which lies adjacent to and directly in front of the stellar object, by more extended redshifted gas situated in the foreground.
The central region of the first-order map appears blueshifted because the blueshifted emission, arising from gas close to and mainly behind the central stellar object, is stronger than the redishifted emission from gas farther away and in front of the stellar object.
This asymmetry is produced when the optical depth is high enough so that at a given line-of-sight velocity the gas facing the observer hides the emission from the gas behind it \citep{1987A&A...186..280A,1991A&A...252..639A}.

		At larger distances from the center, the integrated intensity decreases, the blue and redshifted intensities become similar,
		and the intensity-weighted mean velocity approaches the systemic velocity of the cloud. Therefore, the first-order moment
		of an infalling envelope is characterized by a compact spot of blueshifted emission toward the position of the zeroth-order
		moment peak.
		
		In order to determine the infall velocity, central mass and infall radius we use the hallmark model of \cite{2019A&A...626A..84E}.
		The value of the first-order moment as a function of the angular distance was obtained for the unblended K components of the J=14--13 transition of \ch\, by averaging the first-order moment in concentric rings of width 0\ffas05
		centered on the average position of the peak of the "blue spot", (RA, Dec) (J2000) = (16$^h$56$^m$47.6{$^s$}, --40$^\circ$14$^\prime$25\ffas98). The first-order moment profiles of the different components are presented in Figure \ref{fig:infal}. They seem to belong to two separate groups, with the first-order moment values for the K=2, 3, 4 components being higher than those of the K=6, 7, 8 components, especially near the peak position. However, the K=7 component follows the K=2, 3, 4 components beyond 0\ffas5 distance from the peak.
The difference is likely due to the higher K lines probing the hotter gas close to the HC H {\sc ii} region.
 The best fit is obtained for an infall radius much larger than the beam size, an ambient gas velocity of 
 $V_\mathrm{LSR}$=--12.49$\pm$0.16 \kms, and a central mass of 172.8$\pm8.8 M_{\odot}$ (for a distance of 2.38 kpc and the error does not include the distance uncertainty). 
    To ensure accuracy and assess sensitivity of the model to distance variations, we also fit this model using the near distance of 1.7 kpc, and we get an ambient gas velocity of  $V_\mathrm{LSR}$=--12.46$\pm$0.16 \kms, and a central mass of 126.0$\pm8.7 M_{\odot}$.
This central mass is much larger than that derived from the number of Lyman photons presented in Section \ref{sec:H29a}. This difference could be attributed to a forming very massive star, which is still undergoing infall, emitting fewer ionizing photons than a main sequence star of the same mass.
		The clear detection of the "central blue spot" signature in the G345.0061+01.794B HC H {\sc ii} region
		indicates that infall motions play a fundamental role in the gas kinematics of this source.
  
\subsection{Butterfly pattern}
 In the channel maps (in Figure \ref{fig:channel}) a "butterfly pattern" like structure is observed. The body of the butterfly, a
potentially inflated rotating torus, showing redshifted emission in the east and blueshifted one in the west, extends across the center of the source. Roughly parallel to the rotation axis of this torus we find ionized gas on both sides, i.e. in the
north and south. Extent and thickness of the torus are similar and the ionized gas in the north is oriented slightly eastwards, while that one in the south is also oriented
somewhat eastwards. Hence the butterfly pattern is not perfectly fulfilled, but the observational data come close to it.
Apparently, we see the source from a suitable viewing angle, i.e. the torus is seen approximately edge-on.

\section{Conclusion}

We carried out high angular resolution observations, using ALMA, of emission in highly excited molecular lines of \ch\ and \so\ and in the H29$\alpha$ radio recombination line towards the G345.0061+01.794 B HC H {\sc ii} region. The main results and conclusions are summarized as follows:
\begin{enumerate}
\item Emission was detected in all ten observed K components of the J=14$\rightarrow$13 rotational ladder of \ch\, and in the $30_{4,26}-30_{3,27}$  and $32_{4,28}-32_{3,29}$ lines of \so. The peak of the velocity integrated molecular line intensity is located slightly NW (about 0\ffas4$\pm$0\ffas1) of the peak of the continuum emission. 
\item The first-order moment images of the molecular emission show a central spot of blueshifted emission, with respect to the systemic velocity of the cloud, located at the peak of the zero-order moment, seen in all K-components of \ch\,, in the \so\, lines and in 
        transitions of other species serendipitously detected in the 
        measured frequency band.
\item Rotational diagrams of the emission of the methyl cyanide lines show that the rotational temperature has a peak value of 
252$\pm24$ Kelvin at the position of the "blue spot" and decreases outwards reaching a value of 166$\pm16$ Kelvin at the edge ($\sim$1\arcsec, 0.01 pc for a distance of 2.38 kpc) of the molecular structure, indicating that our observations are probing a hot molecular core that is internally excited. 
In addition, from the emission of four low excitation lines of \so\, we estimate a rotational temperature of 40$\pm$6 Kelvin for the envelope integrated  over the whole source.
\item The first-order moment images and channel maps of the molecular emission show a velocity gradient from roughly east to west with average velocities preferentially blueshifted on the western and redshifted on the eastern side. The change in velocity amounts to 4.3 \kms\ over a region of 3\ffas8 (equivalent to 98 \kms\ pc$^{-1}$ at the distance of 2.38 kpc). 
\item Emission was detected in the H29$\alpha$ line, having a line center velocity of $V_\mathrm{LSR}$=--18.1$\pm$0.9 \kms\,and a linewidth (FWHM) of 33.7$\pm$2.3 \kms. The position of the peak in the velocity integrated emission is coincident with that of the dust continuum. The radio recombination line observations indicate that the ionized gas emission arises from a region having a radius of 0.0037 pc, a mass of ionized gas of (3.6$\pm$0.3)$\times$10$^{-3}$ $M_{\odot}$,  an electron temperature of  8094$\pm$1534 Kelvin, an emission measure of (4.6$\pm$0.9)$\times$10$^8$ pc cm$^{-6}$, and an electron density (2.1$\pm$0.2$)\times$10$^5$ cm$^{-3}$ (for a distance of 2.38 kpc).
\item We modeled  the kinematical characteristics of a "central blue spot" feature as due to infalling motions, deriving a central mass of 172.8$\pm8.8 M_{\odot}$ (for a distance of 2.38 kpc). 
We conclude that the HC H {\sc ii} region is surrounded by a compact structure of hot molecular gas, which is rotating and infalling toward a central mass of 172.8$\pm8.8 M_{\odot}$, that is most likely confining the region of ionized gas. 
\item The properties for the source are reminiscent of the theoretically
proposed "butterfly pattern", with the rotating torus seen
almost edge-on and the ionized gas extending roughly perpendicular
to it.
\end{enumerate}

\begin{acknowledgements}
The authors would like to express their sincere gratitude to Eric Keto for his critical evaluation of the paper.      
      This research was funded by the Science Committee of the Ministry of Science and Higher Education of the Republic of Kazakhstan (Grant Nos. AP13067768 and AP14870504) and sponsored (in part) by the Chinese Academy of Sciences (CAS), through a grant to the CAS South America Center for Astronomy (CASSACA) in Santiago, Chile. GG acknowledges support from ANID BASAL project FB210003. 
RE acknowledges partial financial support from the grants PID2020-117710GB-I00 and CEX2019-000918-M funded by MCIN/ AEI /10.13039/501100011033.
JE acknowledges support from the National Key R\&D Program of China under grant No.2022YFA1603103 and the Regional Collaborative Innovation Project of Xinjiang Uyghur Autonomous Region grant 2022E01050.
DL acknowledges support from National Natural Science Foundation of China (NSFC) through grant No. 12173075 and support from Youth Innovation Promotion Association CAS.
YH acknowledges support from the CAS "Light of West China" Program under grant No. 2020-XBQNXZ-017 and the Xinjiang Key Laboratory of Radio Astrophysics under grant No. 2023D04033.
This paper makes use of the following ALMA data: ADS/JAO.ALMA\#2015.1.01371.S. ALMA is a partnership of ESO (representing its member states), NSF (USA) and NINS (Japan), together with NRC (Canada), MOST and ASIAA (Taiwan), and KASI (Republic of Korea), in cooperation with the Republic of Chile. The Joint ALMA Observatory is operated by ESO, AUI/NRAO and NAOJ.
\end{acknowledgements}

%% To help institutions obtain information on the effectiveness of their 
%% telescopes the AAS Journals has created a group of keywords for telescope 
%% facilities.
%
%% Following the acknowledgments section, use the following syntax and the
%% \facility{} or \facilities{} macros to list the keywords of facilities used 
%% in the research for the paper.  Each keyword is check against the master 
%% list during copy editing.  Individual instruments can be provided in 
%% parentheses, after the keyword, but they are not verified.

\vspace{5mm}
\facilities{ALMA}

%% Similar to \facility{}, there is the optional \software command to allow 
%% authors a place to specify which programs were used during the creation of 
%% the manuscript. Authors should list each code and include either a
%% citation or url to the code inside ()s when available.

\software{astropy \citep{2013A&A...558A..33A,2018AJ....156..123A},  
          CASA \citep{2007ASPC..376..127M},
          MADCUBA \citep{2019A&A...631A.159M}}

%% Appendix material should be preceded with a single \appendix command.
%% There should be a \section command for each appendix. Mark appendix
%% subsections with the same markup you use in the main body of the paper.

%% Each Appendix (indicated with \section) will be lettered A, B, C, etc.
%% The equation counter will reset when it encounters the \appendix
%% command and will number appendix equations (A1), (A2), etc. The
%% Figure and Table counter will not reset.

%% For this sample we use BibTeX plus aasjournals.bst to generate the
%% the bibliography. The sample631.bib file was populated from ADS. To
%% get the citations to show in the compiled file do the following:
%%
%% pdflatex sample631.tex
%% bibtext sample631
%% pdflatex sample631.tex
%% pdflatex sample631.tex

\bibliographystyle{aasjournal}
\bibliography{ALMA_345}

\begin{thebibliography}{}
\expandafter\ifx\csname natexlab\endcsname\relax\def\natexlab#1{#1}\fi
\providecommand{\url}[1]{\href{#1}{#1}}
\providecommand{\dodoi}[1]{doi:~\href{http://doi.org/#1}{\nolinkurl{#1}}}
\providecommand{\doeprint}[1]{\href{http://ascl.net/#1}{\nolinkurl{http://ascl.net/#1}}}
\providecommand{\doarXiv}[1]{\href{https://arxiv.org/abs/#1}{\nolinkurl{https://arxiv.org/abs/#1}}}

\bibitem[{{Anglada} {et~al.}(1991){Anglada}, {Estalella}, {Rodriguez}, \& {Lopez}}]{1991A&A...252..639A}
{Anglada}, G., {Estalella}, R., {Rodriguez}, L.~F., \& {Lopez}, J. C.~R. 1991, \aap, 252, 639

\bibitem[{{Anglada} {et~al.}(1987){Anglada}, {Rodriguez}, {Canto}, {Estalella}, \& {Lopez}}]{1987A&A...186..280A}
{Anglada}, G., {Rodriguez}, L.~F., {Canto}, J., {Estalella}, R., \& {Lopez}, R. 1987, \aap, 186, 280

\bibitem[{{Astropy Collaboration} {et~al.}(2013){Astropy Collaboration}, {Robitaille}, {Tollerud}, {Greenfield}, {Droettboom}, {Bray}, {Aldcroft}, {Davis}, {Ginsburg}, {Price-Whelan}, {Kerzendorf}, {Conley}, {Crighton}, {Barbary}, {Muna}, {Ferguson}, {Grollier}, {Parikh}, {Nair}, {Unther}, {Deil}, {Woillez}, {Conseil}, {Kramer}, {Turner}, {Singer}, {Fox}, {Weaver}, {Zabalza}, {Edwards}, {Azalee Bostroem}, {Burke}, {Casey}, {Crawford}, {Dencheva}, {Ely}, {Jenness}, {Labrie}, {Lim}, {Pierfederici}, {Pontzen}, {Ptak}, {Refsdal}, {Servillat}, \& {Streicher}}]{2013A&A...558A..33A}
{Astropy Collaboration}, {Robitaille}, T.~P., {Tollerud}, E.~J., {et~al.} 2013, \aap, 558, A33, \dodoi{10.1051/0004-6361/201322068}

\bibitem[{{Astropy Collaboration} {et~al.}(2018){Astropy Collaboration}, {Price-Whelan}, {Sip{\H{o}}cz}, {G{\"u}nther}, {Lim}, {Crawford}, {Conseil}, {Shupe}, {Craig}, {Dencheva}, {Ginsburg}, {VanderPlas}, {Bradley}, {P{\'e}rez-Su{\'a}rez}, {de Val-Borro}, {Aldcroft}, {Cruz}, {Robitaille}, {Tollerud}, {Ardelean}, {Babej}, {Bach}, {Bachetti}, {Bakanov}, {Bamford}, {Barentsen}, {Barmby}, {Baumbach}, {Berry}, {Biscani}, {Boquien}, {Bostroem}, {Bouma}, {Brammer}, {Bray}, {Breytenbach}, {Buddelmeijer}, {Burke}, {Calderone}, {Cano Rodr{\'\i}guez}, {Cara}, {Cardoso}, {Cheedella}, {Copin}, {Corrales}, {Crichton}, {D'Avella}, {Deil}, {Depagne}, {Dietrich}, {Donath}, {Droettboom}, {Earl}, {Erben}, {Fabbro}, {Ferreira}, {Finethy}, {Fox}, {Garrison}, {Gibbons}, {Goldstein}, {Gommers}, {Greco}, {Greenfield}, {Groener}, {Grollier}, {Hagen}, {Hirst}, {Homeier}, {Horton}, {Hosseinzadeh}, {Hu}, {Hunkeler}, {Ivezi{\'c}}, {Jain}, {Jenness}, {Kanarek}, {Kendrew}, {Kern}, {Kerzendorf}, {Khvalko}, {King}, {Kirkby}, {Kulkarni},
  {Kumar}, {Lee}, {Lenz}, {Littlefair}, {Ma}, {Macleod}, {Mastropietro}, {McCully}, {Montagnac}, {Morris}, {Mueller}, {Mumford}, {Muna}, {Murphy}, {Nelson}, {Nguyen}, {Ninan}, {N{\"o}the}, {Ogaz}, {Oh}, {Parejko}, {Parley}, {Pascual}, {Patil}, {Patil}, {Plunkett}, {Prochaska}, {Rastogi}, {Reddy Janga}, {Sabater}, {Sakurikar}, {Seifert}, {Sherbert}, {Sherwood-Taylor}, {Shih}, {Sick}, {Silbiger}, {Singanamalla}, {Singer}, {Sladen}, {Sooley}, {Sornarajah}, {Streicher}, {Teuben}, {Thomas}, {Tremblay}, {Turner}, {Terr{\'o}n}, {van Kerkwijk}, {de la Vega}, {Watkins}, {Weaver}, {Whitmore}, {Woillez}, {Zabalza}, \& {Astropy Contributors}}]{2018AJ....156..123A}
{Astropy Collaboration}, {Price-Whelan}, A.~M., {Sip{\H{o}}cz}, B.~M., {et~al.} 2018, \aj, 156, 123, \dodoi{10.3847/1538-3881/aabc4f}

\bibitem[{{Beltr{\'a}n} {et~al.}(2014){Beltr{\'a}n}, {S{\'a}nchez-Monge}, {Cesaroni}, {Kumar}, {Galli}, {Walmsley}, {Etoka}, {Furuya}, {Moscadelli}, {Stanke}, {van der Tak}, {Vig}, {Wang}, {Zinnecker}, {Elia}, \& {Schisano}}]{2014A&A...571A..52B}
{Beltr{\'a}n}, M.~T., {S{\'a}nchez-Monge}, {\'A}., {Cesaroni}, R., {et~al.} 2014, \aap, 571, A52, \dodoi{10.1051/0004-6361/201424031}

\bibitem[{{Estalella} {et~al.}(2019){Estalella}, {Anglada}, {D{\'\i}az-Rodr{\'\i}guez}, \& {Mayen-Gijon}}]{2019A&A...626A..84E}
{Estalella}, R., {Anglada}, G., {D{\'\i}az-Rodr{\'\i}guez}, A.~K., \& {Mayen-Gijon}, J.~M. 2019, \aap, 626, A84, \dodoi{10.1051/0004-6361/201834998}

\bibitem[{{Garay} \& {Lizano}(1999)}]{1999PASP..111.1049G}
{Garay}, G., \& {Lizano}, S. 1999, \pasp, 111, 1049, \dodoi{10.1086/316416}

\bibitem[{{Gordon} \& {Sorochenko}(2002)}]{2002ASSL..282.....G}
{Gordon}, M.~A., \& {Sorochenko}, R.~L. 2002, {Radio Recombination Lines. Their Physics and Astronomical Applications}, \dodoi{10.1007/978-0-387-09604-9}

\bibitem[{{Guzm{\'a}n} {et~al.}(2012){Guzm{\'a}n}, {Garay}, {Brooks}, \& {Voronkov}}]{2012ApJ...753...51G}
{Guzm{\'a}n}, A.~E., {Garay}, G., {Brooks}, K.~J., \& {Voronkov}, M.~A. 2012, \apj, 753, 51, \dodoi{10.1088/0004-637X/753/1/51}

\bibitem[{{Guzm{\'a}n} {et~al.}(2020){Guzm{\'a}n}, {Sanhueza}, {Zapata}, {Garay}, \& {Rodr{\'\i}guez}}]{2020ApJ...904...77G}
{Guzm{\'a}n}, A.~E., {Sanhueza}, P., {Zapata}, L., {Garay}, G., \& {Rodr{\'\i}guez}, L.~F. 2020, \apj, 904, 77, \dodoi{10.3847/1538-4357/abbe09}

\bibitem[{{Guzm{\'a}n} {et~al.}(2014){Guzm{\'a}n}, {Garay}, {Rodr{\'\i}guez}, {Moran}, {Brooks}, {Bronfman}, {Nyman}, {Sanhueza}, \& {Mardones}}]{2014ApJ...796..117G}
{Guzm{\'a}n}, A.~E., {Garay}, G., {Rodr{\'\i}guez}, L.~F., {et~al.} 2014, \apj, 796, 117, \dodoi{10.1088/0004-637X/796/2/117}

\bibitem[{{Hosokawa} \& {Omukai}(2009)}]{2009ApJ...691..823H}
{Hosokawa}, T., \& {Omukai}, K. 2009, \apj, 691, 823, \dodoi{10.1088/0004-637X/691/1/823}

\bibitem[{{Keto}(2007)}]{2007ApJ...666..976K}
{Keto}, E. 2007, \apj, 666, 976, \dodoi{10.1086/520320}

\bibitem[{{Keto} \& {Wood}(2006)}]{2006ApJ...637..850K}
{Keto}, E., \& {Wood}, K. 2006, \apj, 637, 850, \dodoi{10.1086/498611}

\bibitem[{{Kuiper} {et~al.}(2011){Kuiper}, {Klahr}, {Beuther}, \& {Henning}}]{2011ApJ...732...20K}
{Kuiper}, R., {Klahr}, H., {Beuther}, H., \& {Henning}, T. 2011, \apj, 732, 20, \dodoi{10.1088/0004-637X/732/1/20}

\bibitem[{{Kurtz} {et~al.}(2000){Kurtz}, {Cesaroni}, {Churchwell}, {Hofner}, \& {Walmsley}}]{2000prpl.conf..299K}
{Kurtz}, S., {Cesaroni}, R., {Churchwell}, E., {Hofner}, P., \& {Walmsley}, C.~M. 2000, in Protostars and Planets IV, ed. V.~{Mannings}, A.~P. {Boss}, \& S.~S. {Russell}, 299--326

\bibitem[{{Kurtz}(2000)}]{2000RMxAC...9..169K}
{Kurtz}, S.~E. 2000, in Revista Mexicana de Astronomia y Astrofisica Conference Series, Vol.~9, Revista Mexicana de Astronomia y Astrofisica Conference Series, ed. S.~J. {Arthur}, N.~S. {Brickhouse}, \& J.~{Franco}, 169--176

\bibitem[{{Mart{\'\i}n} {et~al.}(2019){Mart{\'\i}n}, {Mart{\'\i}n-Pintado}, {Blanco-S{\'a}nchez}, {Rivilla}, {Rodr{\'\i}guez-Franco}, \& {Rico-Villas}}]{2019A&A...631A.159M}
{Mart{\'\i}n}, S., {Mart{\'\i}n-Pintado}, J., {Blanco-S{\'a}nchez}, C., {et~al.} 2019, \aap, 631, A159, \dodoi{10.1051/0004-6361/201936144}

\bibitem[{{Mayen-Gijon} {et~al.}(2014){Mayen-Gijon}, {Anglada}, {Osorio}, {Rodr{\'\i}guez}, {Lizano}, {G{\'o}mez}, \& {Carrasco-Gonz{\'a}lez}}]{2014MNRAS.437.3766M}
{Mayen-Gijon}, J.~M., {Anglada}, G., {Osorio}, M., {et~al.} 2014, \mnras, 437, 3766, \dodoi{10.1093/mnras/stt2172}

\bibitem[{{McMullin} {et~al.}(2007){McMullin}, {Waters}, {Schiebel}, {Young}, \& {Golap}}]{2007ASPC..376..127M}
{McMullin}, J.~P., {Waters}, B., {Schiebel}, D., {Young}, W., \& {Golap}, K. 2007, in Astronomical Society of the Pacific Conference Series, Vol. 376, Astronomical Data Analysis Software and Systems XVI, ed. R.~A. {Shaw}, F.~{Hill}, \& D.~J. {Bell}, 127

\bibitem[{{Mehringer}(1994)}]{1994ApJS...91..713M}
{Mehringer}, D.~M. 1994, \apjs, 91, 713, \dodoi{10.1086/191953}

\bibitem[{{Mezger} \& {Henderson}(1967)}]{1967ApJ...147..471M}
{Mezger}, P.~G., \& {Henderson}, A.~P. 1967, \apj, 147, 471, \dodoi{10.1086/149030}

\bibitem[{{Mezger} {et~al.}(1974){Mezger}, {Smith}, \& {Churchwell}}]{1974A&A....32..269M}
{Mezger}, P.~G., {Smith}, L.~F., \& {Churchwell}, E. 1974, \aap, 32, 269

\bibitem[{{Mois{\'e}s} {et~al.}(2011){Mois{\'e}s}, {Damineli}, {Figuer{\^e}do}, {Blum}, {Conti}, \& {Barbosa}}]{2011MNRAS.411..705M}
{Mois{\'e}s}, A.~P., {Damineli}, A., {Figuer{\^e}do}, E., {et~al.} 2011, \mnras, 411, 705, \dodoi{10.1111/j.1365-2966.2010.17713.x}

\bibitem[{{Nakano}(1989)}]{1989ApJ...345..464N}
{Nakano}, T. 1989, \apj, 345, 464, \dodoi{10.1086/167919}

\bibitem[{{Narayanan} {et~al.}(1998){Narayanan}, {Walker}, \& {Buckley}}]{1998ApJ...496..292N}
{Narayanan}, G., {Walker}, C.~K., \& {Buckley}, H.~D. 1998, \apj, 496, 292, \dodoi{10.1086/305363}

\bibitem[{{Norberg} \& {Maeder}(2000)}]{2000A&A...359.1025N}
{Norberg}, P., \& {Maeder}, A. 2000, \aap, 359, 1025.
\newblock \doarXiv{astro-ph/0005535}

\bibitem[{{Ossenkopf} \& {Henning}(1994)}]{1994A&A...291..943O}
{Ossenkopf}, V., \& {Henning}, T. 1994, \aap, 291, 943

\bibitem[{{Panagia}(1973)}]{1973AJ.....78..929P}
{Panagia}, N. 1973, \aj, 78, 929, \dodoi{10.1086/111498}

\bibitem[{{S{\'a}nchez-Monge} {et~al.}(2013){S{\'a}nchez-Monge}, {Cesaroni}, {Beltr{\'a}n}, {Kumar}, {Stanke}, {Zinnecker}, {Etoka}, {Galli}, {Hummel}, {Moscadelli}, {Preibisch}, {Ratzka}, {van der Tak}, {Vig}, {Walmsley}, \& {Wang}}]{2013A&A...552L..10S}
{S{\'a}nchez-Monge}, {\'A}., {Cesaroni}, R., {Beltr{\'a}n}, M.~T., {et~al.} 2013, \aap, 552, L10, \dodoi{10.1051/0004-6361/201321134}

\bibitem[{{Scoville} \& {Kwan}(1976)}]{1976ApJ...206..718S}
{Scoville}, N.~Z., \& {Kwan}, J. 1976, \apj, 206, 718, \dodoi{10.1086/154432}

\bibitem[{{Tan} {et~al.}(2014){Tan}, {Beltr{\'a}n}, {Caselli}, {Fontani}, {Fuente}, {Krumholz}, {McKee}, \& {Stolte}}]{2014prpl.conf..149T}
{Tan}, J.~C., {Beltr{\'a}n}, M.~T., {Caselli}, P., {et~al.} 2014, in Protostars and Planets VI, ed. H.~{Beuther}, R.~S. {Klessen}, C.~P. {Dullemond}, \& T.~{Henning}, 149, \dodoi{10.2458/azu\_uapress\_9780816531240-ch007}

\bibitem[{{Urquhart} {et~al.}(2007){Urquhart}, {Busfield}, {Hoare}, {Lumsden}, {Oudmaijer}, {Moore}, {Gibb}, {Purcell}, {Burton}, \& {Marechal}}]{2007A&A...474..891U}
{Urquhart}, J.~S., {Busfield}, A.~L., {Hoare}, M.~G., {et~al.} 2007, \aap, 474, 891, \dodoi{10.1051/0004-6361:20078025}

\bibitem[{{Zapata} {et~al.}(2015){Zapata}, {Palau}, {Galv{\'a}n-Madrid}, {Rodr{\'\i}guez}, {Garay}, {Moran}, \& {Franco-Hern{\'a}ndez}}]{2015MNRAS.447.1826Z}
{Zapata}, L.~A., {Palau}, A., {Galv{\'a}n-Madrid}, R., {et~al.} 2015, \mnras, 447, 1826, \dodoi{10.1093/mnras/stu2527}

\bibitem[{{Zhang} {et~al.}(2014){Zhang}, {Tan}, \& {Hosokawa}}]{2014ApJ...788..166Z}
{Zhang}, Y., {Tan}, J.~C., \& {Hosokawa}, T. 2014, \apj, 788, 166, \dodoi{10.1088/0004-637X/788/2/166}

\end{thebibliography}
%% This command is needed to show the entire author+affiliation list when
%% the collaboration and author truncation commands are used.  It has to
%% go at the end of the manuscript.
%\allauthors

%% Include this line if you are using the \added, \replaced, \deleted
%% commands to see a summary list of all changes at the end of the article.
%\listofchanges

\end{document}